\documentclass[aps,superscriptaddress,twocolumn,pra,draft,showpacs]{revtex4}
\usepackage{amsmath}
  
\newcommand*{\Eq}[1]{Eq.~(\ref{#1})}
\newcommand*{\ket}[1]{|#1\rangle}
\newcommand*{\bra}[1]{\langle#1|}

\begin{document}


\title{Ancilla-Assisted Enhancement of Channel Estimation for 
Low-Noise Parameters}


\author{Masahiro Hotta}
\email[]{hotta@tuhep.phys.tohoku.ac.jp}
\affiliation{Department of Physics, Faculty of Science, T\^{o}hoku University,
Aoba-ku, Sendai, 980-8578, Japan}

\author{Tokishiro Karasawa}
\email[]{jidai@ims.is.tohoku.ac.jp}
\affiliation{Graduate School of Information Sciences,
T\^{o}hoku University, Aoba-ku, Sendai,  980-8579, Japan}

\author{Masanao Ozawa}
\email[]{ozawa@math.is.tohoku.ac.jp}
\affiliation{Graduate School of Information Sciences,
T\^{o}hoku University, Aoba-ku, Sendai,  980-8579, Japan}


\date{Revised August 29, 2005}
\begin{abstract}
In order to make a unified treatment for estimation problems 
of a very small noise or a very weak signal in a quantum process,  
we introduce the notion of a low-noise quantum channel with one noise parameter.
It is known in several examples that prior entanglement together 
with nonlocal output measurement improves the performance of  the channel estimation.  
In this paper, we study this ``ancilla-assisted enhancement''
for estimation of the noise parameter in a general low-noise channel.
For channels on two level systems we
prove that the enhancement factor, 
the ratio of the Fisher information of the ancilla-assisted estimation 
to that of the original one, is always upper bounded by $3/2$. 
Some conditions for the attainability are also
given with illustrative examples.
\end{abstract}


\maketitle

\section{Introduction}

One of the formidable obstacles for the realization of quantum computers is
decoherence caused by the coupling between computational qubits and the environment.
Recent study of quantum error correction has shown that fault-tolerant
quantum computing is in principle possible, but it requires that the noise
caused by the decoherence should be lower than the very stringent threshold.
Obviously, such a statement has a physical meaning only if we have an efficient
method for quantitatively estimating very small noise in quantum devices in
real experiments.  However,
if the noise is very small, so is our success probability of observing the
disturbance caused by that noise.  This difficulty makes evident the demand
for the study of optimal quantum estimation of very small nose in general
quantum channels based on well-established quantum estimation theory.

Quantum estimation theory was instituted by Helstrom in the late 1960's
and has been developed with various applications until recently; 
for standard reviews we refer to Helstrom \cite{Hel76} and  Holevo \cite{Hol82}, 
and see also Hayashi \cite{Hay05} for recent progress.  A typical problem of quantum
estimation is to ask what is the best observable, possibly in an extended system with
ancilla, to measure in order to estimate the true value of
$\theta$ provided that the system is known to be in one of the state in a given
family $\{\rho_{\theta}\}$.  
A well-established solution for this problem
is given as follows.  
We call an observable $A$ a (locally) unbiased estimator at $\theta=\theta_0$ 
if the expectation value $E_{\theta}[A]$ of $A$ in the state $\rho_{\theta}$ satisfies
\begin{eqnarray}
E_{\theta_0}[A]&=&\theta_0,\\
\partial_{\theta}E_{\theta}[A]|_{\theta=\theta_0}&=&1.
\end{eqnarray} 
In general there are many unbiased estimators.
In order to select a good one, we consider the variance
$V_\theta [A]$ of an arbitrary unbiased estimator $A$ 
in the state $\rho_{\theta}$.
Then, the quantum Cram\'{e}r-Rao inequality 
\begin{equation}\label{CR}
V_\theta [A]\ge \frac{1}{J(\rho_\theta)}
\end{equation}
holds for any unbiased estimator $A$ at $\theta$, 
where 
\begin{equation} \label{FI}
J(\rho_\theta)={\rm Tr}[\rho_{\theta}L^{2}_{\theta}]
\end{equation} 
is the (quantum) Fisher information defined through the
symmetric logarithmic derivative (SLD) $L_{\theta}$
that is characterized by the relations
\begin{eqnarray}
\partial_\theta \rho_\theta
 &=& \frac{1}{2}(L_\theta \rho_\theta +\rho_\theta L_\theta),
\label{SLD1}\\
L_{\theta}^{\dagger}&=&L_{\theta}.
\label{SLD2}
\end{eqnarray}
The SLD is determined uniquely on the range of $\rho_{\theta}$, i.e., 
$L_{\theta}\rho_{\theta}=L'_{\theta}\rho_{\theta}$ holds for any two SLDs
$L_{\theta}$ and $L'_{\theta}$.
The Cram\'{e}r-Rao inequality (\ref{CR}) follows from a simple application of 
the Schwarz inequality for the Hilbert-Schmidt inner product.
From the equality condition for that the lower bound $J_\theta^{-1}$ in Eq.~(\ref{CR})
is always achieved by any observable $A$ satisfying
\begin{equation}\label{Optimality}
A\rho_{\theta}=(J_{\theta}^{-1}L_{\theta}+\theta)\rho_{\theta},
\end{equation}
see Refs.~\cite{Hel76,Hol82} and for a straightforward derivation see
Appendix of Ref.~\cite{04QEL}.
In general, to find an optimal estimator for the true value $\theta$ needs prior information
on the value $\theta$, which might be collected by prior estimations assuming prior probabilities
on the unknown parameter, so that the optimal estimator is considered as an ultimate limit
allowed by physics.  However, there are some cases in which the optimal estimator can 
be chosen uniformly over unknown values of $\theta$ \cite{FN95}.  In these cases the
ultimate limit can be certainly achieved without prior information.

From the quantum estimation theory for state parameters mentioned above,
we can construct an estimation theory for unknown parameters of 
physical processes, such as coupling constants of the interaction.
Suppose that we prepare a quantum system in an initial state $\rho_{in}$ and
leave it in an evolution process characterized by an unknown parameter $\theta$.  
Then, the final state $\rho_{out}(\theta)$ of this process depends on the parameter $\theta$.  
The problem of finding the optimal estimation of the parameter $\theta$ 
is solved by maximizing the Fisher information  $J_{\theta}$ over all
the possible initial states $\rho_{in}$ and all the
possible observable $A$ in the final state  \cite{CN97,PCZ97}.  
The above physical process can be represented by a mapping 
$\Gamma_\theta$
that transform the
initial state $\rho_{in}$ to the final state $\rho_{out}$ as
\begin{equation}
\rho_{out} =\Gamma_\theta [\rho_{in}].
\end{equation}
It is now fairly well-known that every general state change,
called a quantum operation or a quantum channel, such as $\Gamma_\theta$,
physically realizable with probability one should be a trace-preserving 
completely positive (TPCP) mapping,  and conversely that every TPCP map 
can be realized as a unitary process of the system augmented 
by an ancilla prepared in a fixed state as shown by Kraus \cite{Kra71,Kra83};
see also Ref.~\cite{83CR,84QC} for the generalization of the above statement 
to generalized measurements and see Ref.~\cite{04URN} for the latest elaboration.

As pointed out in Ref.~\cite{Fuj01}, one can improve the parameter estimation
if  a correlation, or in particular an entanglement, is allowed between the
input system $S$ and an ancilla $A$.  It should be stressed that in doing
so one needs no physical process to occur on the ancilla system $A$ while
the system $S$ passes through the channel $\Gamma_{\theta}$.  
In this case,  the extended channel is represented as
$\Gamma_\theta \otimes id_{A}$, where $id_A$ stands for the identity channel
for $A$.   Then, the improvement can be achieved by the initial 
preparation of the composite system in an entangled state together with
the measurement of the composite system after the process.

Recent progress has been reported on problems for special families of
quantum channels,  in particular, 
SU(2) channel \cite{Fuj02}, 
a generalized Pauli channel \cite{FI03},  
a generalized amplitude-damping channel \cite{Fuj04},  
U(N) channel and its Abelian subgroup channel \cite{Bal04a,Bal04b}.  
A review by Fujiwara \cite{Fuj04b} is also available. 
For earlier contributions see also \cite{CPR00,Aci01,DPP01,FMCF01,BFF01,SBB02}.
In this paper, we are devoted to the ancilla-assisted enhancement of Fisher information 
derived by the quantum Cram\'{e}r-Rao bound, whereas ancilla-assisted enhancements
have been recently investigated within the Bayesian approach \cite{CDS05,Sac05,Sac05a} 
and the minimax approach \cite{DSK05}.

This enhancement effect not only projects a theoretical profundity of
quantum mechanics, but also suggests many physical applications including
the low-noise estimation  in quantum computing, where the enhanced noise
estimation is expected to contribute to developing the quantum error correction 
and quantum noise reduction technology \cite{NC00}.  

We can find another application of  the low-noise estimation
in elementary particle physics. Today, 
 because of technological difficulties of high-energy experiments,
  direct researches of  new physics  far beyond the TeV energy scale
 are almost impossible. This is  one of reasons why  
 the low-energy rare processes predicted 
  by the new physics  recently attract much attention. 
  (The CPT symmetry violation in the $K-\bar{K}$ oscillation is  one of the typical processes
\cite{Ebe72,CCE76,EHN84,HP95}.)
 Clearly, the number of signals for 
 the new-physics evidence is predicted very small, even if the process really 
 exists in nature. The new-physics data should be separated from  
  an enormous number of ordinary data explained by the standard model. 
  This means that  the 
  new-physics data can be regarded as a 
sort of background low noise in the standard data. 
Hence, 
 we can treat the rare process as a low-noise channel. 
It is very significant to estimate the intensity of the low noise  
 because indirect information about physics 
 beyond the standard model is obtained. In the estimation, 
  the above ancilla-assisted enhancement may effectively reduce the trial 
  number of the experiment.

In this paper, we study the estimation theory of the parameter
characterizing a small noise in a general quantum channel
on a system with finite dimensional state space.
We can always decompose the quantum channel into two channels so that the
input state of the original channel passes through the first noiseless
channel and consecutively passes through the second noisy quantum channel
called the noise channel.
Thus, we can concentrate our attention on the noise channel.
We are interested in the case where the noise is so small that the noise
channel deviates only a little from the identity channel.  In such a case,
the channel is called a low-noise channel, and the parameter representing
the noise is called the low-noise parameter denoted by $\epsilon$.   
Let $\Gamma_\epsilon$ be a low-noise channel with low-noise parameter $\epsilon$.  
We assume that the low-noise parameter is scaled so that $\Gamma_0$ is the identity channel.
We can formulate natural mathematical requirements for the behavior of the
low-noise parameter in a neighborhood of $\epsilon=0$.
It is an interesting problem to figure out how much  ancilla-assisted
enhancement can be achievable in the estimation of the low-noise parameter
$\epsilon$.
In this paper we shall discuss this problem and obtain several upper bounds
for this ancilla-assisted enhancement factor in the low-noise parameter
estimation.

In Section 2, we explain a theorem  \cite{Fuj01} states that
the Fisher information is attained in a pure initial state, so that we can
always assume that the input of the channel is a pure state.
In Section 3, we discuss parameter estimation for unitary channels, which do
not couple with the environment, and show that we have no ancilla-assisted
enhancement.
Thus, the ancilla-assisted enhancement is possible only for channels coupled
with the environment.
In Section 4, we introduce the notion of low-noise channels mentioned above
with rigorous mathematical requirements, and we obtain a general formula for
the upper bound for the ancilla-assisted enhancement factor.
In Section 5, we introduce two physical examples of low-noise channel.
In Section 6, we give a concrete evaluation of the enhancement factor in two
level systems.
Let us consider a low-noise channel  $\Gamma_\epsilon$
with low-noise parameter $\epsilon$ in a two level system $S_2$.  We obtain
a universal upper bound for the enhancement factor $\eta$ defined by
\begin{equation}
\eta =\frac{{\cal L}[\max[J_{S_2 +A}]_{\rho_{S_2 +A}}]}
{{\cal L}[\max[J_{S_2}]_{\rho_{S_2}}]} \label{ed}
\end{equation}
for any finite level ancilla $A$.  
Here, $\rho_{S_2}$ is the input in the system $S$, 
$J_{S_2}$ is the Fisher information of 
$\Gamma_\epsilon [\rho_{S_2}]$, 
$\rho_{S_2 +A}$ is the channel input in the composite system $S_2 +A$,
$J_{S_2 +A}$ is the Fisher information of the  output states
$(\Gamma_\epsilon \otimes id_A) [\rho_{S_2 +A}]$,
and $\max[ \cdot ]_\rho$ stands for the maximum over all the state $\rho$.
As shown later, $J_{S_2}$ and $J_{S_2 +A}$ shows a singular behavior
$\propto 1/\epsilon$ in the $\epsilon$ expansion, and ${\cal L}[J]$ is
coefficient of $\propto 1/\epsilon$, i.e.,
\begin{equation}
J(\epsilon) =\frac{1}{\epsilon}{\cal L}[J] +O(\epsilon^0).
\end{equation} 

The universal upper bound of the enhancement factor $\eta$ for all the two
level systems is given by
\begin{equation}
\eta  \leq \frac{3}{2} .
\end{equation}
This upper bound is attainable by various channels $\Gamma_\epsilon$, and
the corresponding optimal input state is a maximal entangled state, and
holds for any low-noise channels on two level systems.

\section{The Maximum Is Attained by a Pure Input State: 
The Fujiwara theorem}

In this section we briefly review an important theorem 
due to Fujiwara  \cite{Fuj01}: {\em the maximum of the Fisher information 
of output states $\rho_\theta(=\Gamma_\theta [\rho])$ 
over all possible input states $\rho$ is attained by a pure input 
state for an arbitrary fixed channel $\Gamma_\theta $. }
 
To show this following Fujiwara,  
let  $L_{\theta}$ be the SLD defined by
Eqs.~(\ref{SLD1}), (\ref{SLD2})
for the output state $\rho_\theta$. 
Then the Fisher information $J(\rho_{\theta})$ is given by Eq.~(\ref{FI}).
Fujiwara  \cite{Fuj01} showed that 
the Fisher information has a convexity property, i.e.,
\begin{equation}
J(\lambda \sigma_\theta +(1-\lambda) \tau_\theta)
\leq 
\lambda J(\sigma_\theta)
+(1-\lambda)J(\tau_\theta).\label{convex}
\end{equation}
for any $0<\lambda<1$, 
where $\sigma_\theta$ and $\tau_\theta$ are states with
parameter $\theta$. 

To see the above relation, let Hermitian operators 
$L^{\sigma}_\theta$ and $L^{\tau}_\theta$
be the SLDs of $\sigma_\theta$ and $\tau_\theta$, respectively, i.e., 
\begin{eqnarray}
\partial_\theta \sigma_\theta
& =& \frac{1}{2}(L^\sigma_\theta \sigma_\theta +\sigma_\theta L^\sigma_\theta),\\
\partial_\theta \tau_\theta
& =& \frac{1}{2}(L^\tau_\theta \tau_\theta +\tau_\theta L^\tau_\theta).
\end{eqnarray}
Let us consider the tensor product Hilbert space 
${\cal K}={\cal H}\otimes {\bf C}^{2}$,
where ${\cal H}$ is the state space of $S$ and ${\bf C}^{2}$ is a 
2-dimensional state space. 
With fixed basis $\{\ket{0},\ket{1}\}$ of ${\bf C}^{2}$, let
$\tilde{\rho}_{\theta}$ be a density operator on ${\cal K}$ such that
\begin{equation}
\tilde{\rho}_{\theta}
=\lambda\sigma_{\theta}\otimes\ket{0}\bra{0}+(1-\lambda)\tau_{\theta}\otimes\ket{1}\bra{1}.
\end{equation}
Then, it is easy to see that the SLD of $\tilde{\rho}_{\theta}$ is 
$L^{\sigma}_{\theta}\otimes\ket{0}\bra{0}+L^{\tau}\otimes\ket{1}\bra{1}$,
so that the Fisher information of $\tilde{\rho}_{\theta}$ is given by
\begin{eqnarray}
J(\tilde{\rho}_{\theta})
&=&
{\rm Tr}[\tilde{\rho}_\theta
(L^{\sigma}_{\theta}\otimes\ket{0}\bra{0}+L^{\tau}\otimes\ket{1}\bra{1})^{2}]\nonumber\\
&=&
\lambda{\rm Tr}[{\sigma}_\theta (L^{\sigma}_{\theta})^{2}]+
(1-\lambda){\rm Tr}[{\tau}_\theta (L^{\tau}_{\theta})^{2}]\nonumber\\
&=&
\lambda J({\sigma}_\theta)+(1-\lambda)J({\tau}_\theta).
\label{Fujiwara}
\end{eqnarray}
On the other hand, the partial trace of $\tilde{\rho}_{\theta}$ over ${\bf C}^{2}$
is given by
\begin{eqnarray}
{\rm Tr}_{{\bf C}^{2}}[\tilde{\rho}_{\theta}]
=
\lambda\sigma_{\theta}+(1-\lambda)\tau_{\theta}.
\end{eqnarray}
Since the partial trace is a trace-preserving completely positive map,
the monotonicity of the Fisher information 
under trace-preserving completely positive maps \cite{MC90,Pet96,PS96}
concludes
\begin{equation}
J(\lambda \sigma_\theta +(1-\lambda) \tau_\theta)\leq
J(\tilde{\rho}_{\theta}).
\label{Petz}
\end{equation}
Therefore, from Eq.~(\ref{Fujiwara}) and Eq.~(\ref{Petz}) the convexity  
relation (\ref{convex}) follows.

Now suppose that an input state $\bar\rho$ maximizes 
the Fisher information,
i.e.,
\begin{equation}
J(\Gamma_\theta[\bar\rho])=\max[J(\Gamma_\theta[\rho])]_{\rho}.
\end{equation}
Let
\begin{equation}
\bar\rho =\sum_n p_n  |n\rangle\langle n|
\end{equation}
be the spectral decomposition,
where $0< p_n \leq 1$ and $\sum p_n=1$. 
The output state $\bar\rho_\theta$ is given by 
\begin{equation}
\bar\rho_\theta =\Gamma_\theta [\bar\rho]=
\sum_n p_n \Gamma_\theta [ |n\rangle\langle n|].
\end{equation}
By using  relation (\ref{convex}) repeatedly, we have 
\begin{equation}
J(\bar\rho_\theta)
\leq
\sum_n p_n J(\Gamma_\theta[|n\rangle\langle n|]).
\end{equation}
Since $\bar\rho$ maximizes the Fisher information, we also have
\begin{equation}
\sum_n p_n J(\Gamma_\theta[|n\rangle\langle n|])\leq J(\bar\rho_\theta),
\end{equation}
and this concludes the relation
$J(\bar\rho_\theta)=J(\Gamma_\theta[|n\rangle\langle n|])$ for all $n$.
Thus, the maximum of the Fisher information is also 
attained by a pure input state.

From now on, we assume without any loss of generality 
that the input state of the channel is always a pure state 
by virtue of this theorem.

\section{One-Parameter Unitary Channels Have No Enhancement}

Before we go to general analysis of low-noise channels, let us consider
the case where the channel is unitary, or the channel does not interact
with the environment.
Interestingly, the maximization of the output Fisher information 
$J[\rho_\theta]$ with respect to the input $\rho$
 can be explicitly accomplished.  
After the calculation of the maximum, one can notice that  
 the ancilla-assisted enhancement does not take place at all. 
The result makes it clear that, in order to gain the ancilla-assisted 
enhancement for channel 
parameter estimations, the channels must have the effective 
interaction between the system and the environment.

Let $U(\theta)$ be a unitary operator with an unknown parameter $\theta$. 
Then  the output state of the unitary channel determined by $U(\theta)$
for an input state $|\Psi\rangle$ is given by 
\begin{equation}
\rho(\theta) =|\Psi (\theta)\rangle\langle \Psi (\theta)| ,
\end{equation}
where the output state $|\Psi(\theta)\rangle$ is defined by 
\begin{equation}
|\Psi(\theta)\rangle= U(\theta)|\Psi\rangle.
\end{equation}
By introducing the (logarithmic) Hamiltonian operator $H(\theta)$ such that  
\begin{eqnarray}
H(\theta)&=&i(\partial_\theta U(\theta) )U(\theta)^\dagger,\\
H(\theta)^{\dagger}&=&H(\theta),
\end{eqnarray}
and using the result in  Ref.~\cite{FN95},
the Fisher information of the output state is evaluated as
\begin{equation}\label{FisherInformation}
J_S [\rho_\theta]=4V_{\Psi(\theta)}[H(\theta)],
\end{equation}
where $V_{\Psi(\theta)}[H(\theta)]$ is the variance of $H(\theta)$ in
the state $\Psi(\theta)$, i.e., 
\begin{equation}
V_{\Psi(\theta)}[H(\theta)]=
\langle \Psi(\theta) |H(\theta)^2|\Psi(\theta)\rangle
-\langle \Psi(\theta)|H(\theta)|\Psi(\theta )\rangle^2.
\end{equation}
To obtain the maximum of $J_S$, 
let us consider the maximum and minimum of the eigenvalues $E_n$ of $H(\theta)$:
\begin{equation}
E_{\max}(\theta) = \max[E_n ]_n,
\end{equation}
\begin{equation}
E_{\min}(\theta) = \min[E_n]_n.
\end{equation}
Let 
 $|\max(\theta)\rangle$ and $|\min(\theta)\rangle$
 be eigenstates corresponding to 
 $E_{\max}(\theta)$ and $E_{\min}(\theta)$, respectively.
 By a straightforward manipulation, it is easy to see that 
 the maximum of  $J_S$ is taken by a pure input state 
 $|\Phi\rangle =U(\theta)^\dagger |\Phi (\theta)\rangle $, where 
$|\Phi (\theta)\rangle $ is given by 
\begin{equation}
|\Phi(\theta)\rangle =
\frac{1}{\sqrt{2}}
\left[
|\max(\theta)\rangle +|\min(\theta)\rangle
\right].
\end{equation}
For a fixed value of $\theta$, the maximum is given by
\begin{eqnarray}
\max[ J_S ]_{|\Psi\rangle} &=&
 J_S[ |\Phi(\theta)\rangle\langle\Phi(\theta)|]\nonumber\\
&=&(E_{\max} (\theta) -E_{\min} (\theta))^2.\label{JU}
\end{eqnarray}

To obtain the corresponding result for ancilla-assisted 
estimations, let us introduce an ancilla system $A$ 
and the extended channel defined by 
\begin{equation}
|\tilde{\Psi} (\theta)\rangle
=(U(\theta)\otimes {\bf 1}_A ) |\tilde{\Psi}\rangle,
\end{equation}
where $|\tilde{\Psi}\rangle$ is a state of the composite system 
$S+A$ to be put in the extended channel.
For the output state
$\tilde{\rho}(\theta) =|\tilde{\Psi}(\theta)\rangle\langle\tilde{\Psi}
(\theta) |$, the Fisher information is given by
\begin{equation}
J_{S+A}[\tilde{\rho}(\theta)]
=
4V_{\tilde{\Psi}(\theta)}[H(\theta) \otimes {\bf 1}_A],
\end{equation}
where $V_{\tilde{\Psi}(\theta)}[H(\theta) \otimes {\bf 1}_A]$
is the variance of $H(\theta) \otimes {\bf 1}_A$ in the state 
$\tilde{\Psi}(\theta)$.
Note that the maximum and minimum of the eigenvalues of 
$H\otimes {\bf 1}_A$ are taken in the states $|\max (\theta)\rangle |a\rangle$
 $|\min (\theta)\rangle |a\rangle$, respectively, 
with an arbitrary ancilla state  $|a\rangle$, i.e., 
\begin{eqnarray}
H(\theta)\otimes {\bf 1}_A |\max (\theta)\rangle |a\rangle
 &=&E_{\max} (\theta) |\max (\theta)\rangle |a\rangle, \\
H(\theta)\otimes {\bf 1}_A |\min (\theta)\rangle |a\rangle
 &=&E_{\min} (\theta)|\min (\theta)\rangle |a\rangle.
\end{eqnarray}
Hence, the input state given by 
\begin{eqnarray}
|\tilde{\Phi}\rangle
&=&
\frac{1}{\sqrt{2}}
(U(\theta)^\dagger \otimes  {\bf 1}_A)
\left[
|\max(\theta)\rangle |a\rangle +|\min(\theta)\rangle |a\rangle
\right]\nonumber\\
&=&
|\Phi\rangle |a\rangle
\end{eqnarray}
takes the maximum value of $J_{S+A}$, 
which turns out to be the same as that given in (\ref{JU}),
i.e., 
\begin{eqnarray}
\max[ J_{S+A} ]_{|\tilde{\Psi}\rangle} 
&=& J_{S+A}
[ |\tilde{\Phi}(\theta)\rangle\langle\tilde{\Phi}(\theta)|]
\nonumber\\
&=&(E_{\max} (\theta) -E_{\min} (\theta))^2.
\end{eqnarray}
Consequently, no enhancement by the ancilla extension is observed
 in this unitary case, i.e., 
\begin{equation}
\frac{\max[ J_{S+A} ]_{\rho_{S+A}}}{\max[ J_S ]_{\rho_S}}
=
\frac{\max[ J_{S+A} ]_{|\Psi_{S+A} \rangle}}{\max[ J_S ]_{|\Psi_S \rangle}}
=1.
\end{equation}

It should be noted here that the above argument applies only to one-parameter
unitary channels, for which Eq.~(\ref{FisherInformation}) can be applied, 
whereas a generalization to Abelian group parameters may follow.
For multiple phase parameter estimation of unitary channels, 
Ballester \cite{Bal04a,Bal04b} showed, ancilla-assisted enhancement actually takes place,
whereas for commuting phase parameter estimation no enhancement occurs.

Within Bayesian approach, Chiribella, D'Ariano, and Sacchi \cite{CDS05} 
showed that unitary channels 
with non-Abelian group parameter can have ancilla-assisted improvement of a large
class of cost functions.
In this connection, Sacchi \cite{Sac05,Sac05a} gave extensive analysis on the condition
for ancilla-assisted improvement of the error probability for discrimination of Pauli
channels.   

\section{Low-Noise Channels}

In this section, we introduce the notion of a low-noise
channel  $\Gamma_\epsilon$ with unknown parameter $\epsilon$,
which takes only small values $\epsilon\sim 0$, 
by requiring a physically natural
assumption of the channel $\Gamma_\epsilon$ for the parameter values near
$\epsilon =0$. The small parameter $\epsilon$ is assumed to control the low noise
well  enough and is called the low-noise parameter.
   
 As mentioned in the introduction, 
 we will focus on the ancilla extension of the low-noise channel 
 defined by   $\Gamma_\epsilon \otimes id_{A}$.
  The ancilla-assisted enhancement factor $\eta$ is also 
  defined as the ratio of the Fisher information of the ancilla-assisted estimation 
to that of the original one and is analyzed in detail.

The concept of the noise in a quantum process to implement a
target unitary process can be understood
under the following consideration.  Suppose that 
we would like to implement a unitary channel $\Lambda^{(U)}$ for a system
$S$, so that the output state corresponding to an input state $\rho_{in}$ of 
$S$ is designed to be
\begin{equation}
\rho_{out} =\Lambda^{(U)} [\rho_{in}]= U\rho_{in} U^\dagger.
\end{equation}
Without any noise, the unitary operator can be normally implemented as
\begin{equation}\label{eq:unitary}
U=\exp[-it H_S/\hbar ],
\end{equation}
where $t$ is the time interval from input to output, 
and $H_S$ is the Hamiltonian of $S$ under control
 (Fig.~1).

\begin{widetext}

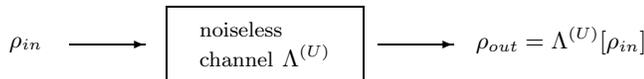
\begin{figure}[hbt]
\begin{center}
\setlength{\unitlength}{1mm}
\begin{picture}(75,10)(0,0)

\put(0,5){\makebox(5,0)[l]{$\rho_{in}$}}

\put(8,5){\vector(1,0){10}}
\put(49,5){\vector(1,0){10}}

\put(21,0){\framebox(26,10)[i]{\shortstack[l]
{{\footnotesize{noiseless}}\\ {\footnotesize{channel}} $\Lambda^{(U)}$ }}}
\put(62,5.5){\makebox(5,0)[l]{$\rho_{out} = \Lambda^{(U)}[ \rho_{in}]$}}
\end{picture}
\end{center}
\caption{{\small
Without any noise, a 
unitary channel  $\Lambda^{(U)}$ can be implemented
with a suitable controlled Hamiltonian $H_{S}$ in a time interval $t$
satisfying eq.~(\ref{eq:unitary}).}
}
\end{figure} 

\end{widetext}

In real life, the system $S$ is coupled weakly with the
environment $E$ and causes the decoherence that cannot be
corrected by controlling the Hamiltonian $H_S$ of the system $S$,
so that the noise is brought from the environment.
Assume that the noise is controlled by one unknown 
positive parameter $\nu$.
  The estimation of the noise parameter 
  $\nu$ often becomes critical in development of
   quantum devices such as quantum computers.

The total Hamiltonian reads 
\begin{equation}
H_{tot} =H_S +H_{SE} +H_E,
\end{equation}
where $H_E$ is the Hamiltonian of $E$ and $H_{SE}$ is the interaction
 Hamiltonian between $S$ and $E$.
 Because of  the noise, 
 the actual output state $\rho_{out}'$ deviates 
 from the intended output state $\rho_{out}$
 (Fig.~2).
\newpage

\begin{widetext} 

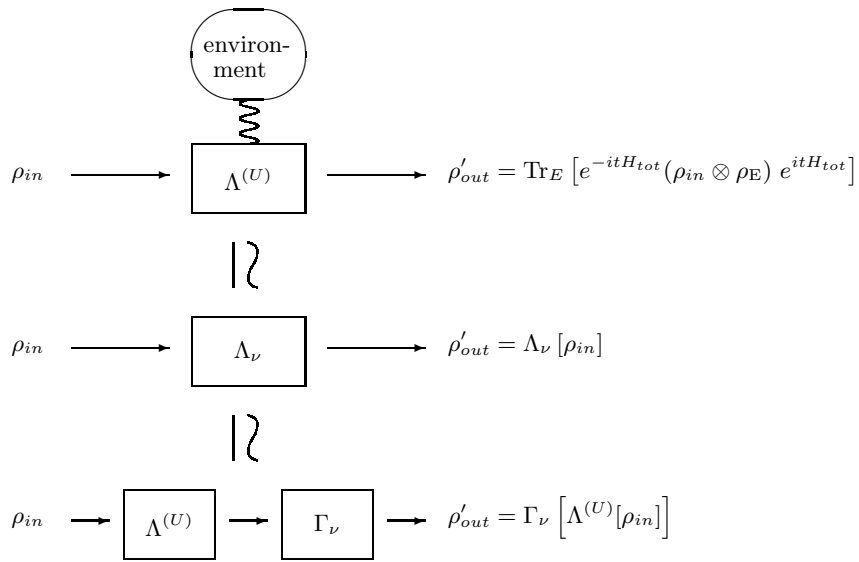
\begin{figure}[hbt]
\begin{center}
\setlength{\unitlength}{1mm}
\begin{picture}(80,70)(0,0)

\put(25.5,68.5){\makebox(5,0)[l]{environ-}}
\put(25.7,65){\makebox(5,0)[l]{   ment}}
\qbezier(31.5,60)(29,60.5)(31.5,61)
\qbezier(31.5,59)(34,59.5)(31.5,60)
\qbezier(31.5,58)(29,58.5)(31.5,59)
\qbezier(31.5,57)(34,57.5)(31.5,58)
\qbezier(31.5,56)(29,56.5)(31.5,57)
\qbezier(31.5,55.1)(34,55.5)(31.5,56)
 \put(31.5,67){{\oval(15.2,12)}}

\put(0,51){\makebox(5,0)[l]{$\rho_{in}$}}
\put(8,51){\vector(1,0){13}}
\put(24,46){\framebox(15,9)[c]{\shortstack[c]{$\Lambda^{(U)}$}}}
\put(42,51){\vector(1,0){13}}
\put(58,51.5){\makebox(5,0)[l]{$\rho_{out}'  ={\mathrm{Tr}}_{E} \left[
e^{-itH_{tot}} (\rho_{in}\otimes \rho_{\mathrm{E}} )\hspace{1mm}e^{itH_{tot}}\right]  $}}

\qbezier(32,39)(33.5,41.5)(31.4,42)
\qbezier(32.6,36)(30.5,36.5)(32,39)
\put(29.5,42){\line(0,-1){6}}

\put(0,28){\makebox(5,0)[l]{$\rho_{in}$}}
\put(8,28){\vector(1,0){13}}
\put(24,23){\framebox(15,9)[c]{\shortstack[c]{$\Lambda_{\nu}$}}}
\put(42,28){\vector(1,0){13}}
\put(58,28.5){\makebox(5,0)[l]{$\rho_{out}' 
= \Lambda_{\nu}\left[ \rho_{in} \right]  $}}

\qbezier(32,16)(33.5,18.5)(31.4,19)
\qbezier(32.6,13)(30.5,13.5)(32,16)
\put(29.5,19){\line(0,-1){6}}

\put(0,5){\makebox(5,0)[l]{$\rho_{in}$}}
\put(8,5){\vector(1,0){5}}
\put(15,0){\framebox(12,9)[c]{\shortstack[c]{$\Lambda^{(U)}$}}}
\put(29,5){\vector(1,0){5}}
\put(36,0){\framebox(12,9)[c]{\shortstack[c]{$\Gamma_{\nu}$}}}
\put(50,5){\vector(1,0){5}}
\put(58,5.5){\makebox(5,0)[l]{$\rho_{out}' 
= \Gamma_{\nu}\left[ \Lambda^{(U)}[\rho_{in}]   \right] $}}
\end{picture}
\end{center}
\caption{
{\small 
A controlled unitary process, such as quantum computing, 
usually suffers from noise (the first line).
The noisy process is described as a TPCP map, a channel $\Lambda_{\nu}$,
parametrized by one noise parameter $\nu$ (the second line).
In quantum theory, the disturbed process is equivalently 
described by a sequence of two channels. 
The first channel is the originally intended unitary channel $\Lambda^{(U)}$.
The second channel  $\Gamma_{\nu}$ describes the genuine noise effect. 
We call  $\Gamma_{\nu}$ the noise channel (the third line).}}
\end{figure} 
\end{widetext}

By using $H_{tot}$, 
 the output state $\rho_{out}'$ is determined in principle by
\begin{eqnarray}
\rho_{out}'
=
{\rm Tr}_E \left[
e^{-itH_{tot}} \left(\rho_{in}\otimes \rho_E\right) e^{itH_{tot}}
\right], 
\label{1ds}
\end{eqnarray}
where ${\rm Tr}_E$ is the partial trace over $E$ and 
$\rho_E$ is the initial state of $E$. 
Theoretically, it is  preferable that 
 we determine the value of the noise parameter $\nu$  
  via \Eq{1ds}; however, the explicit calculation of
  \Eq{1ds}  is  too 
 complicated to perform in many cases. 
Hence, adopting 
 a reasonable theoretical model of the noise effect, 
  the actual value of its noise parameter of the model should be
  experimentally estimated. 

Without assuming any detailed knowledge about $H_E$ and $H_{SE}$,
it is natural to represent the noisy process by  a 
TPCP map $\Lambda_\nu$ such that
\begin{equation}
\rho_{out}' = \Lambda_\nu [\rho_{in} ],
\end{equation}
where the relation $\Lambda_0 =\Lambda^{(U)}$ holds 
as the noiseless case. 
In quantum theory, the channel $\Lambda_\nu$ can be equivalently described 
 by a sequence of two channels (the third line of Fig.~2).  
The first one is the target unitary channel $\Lambda^{(U)}$ 
and the second represents the genuine noise part.
This means that the general noisy process is equivalent to the 
noiseless unitary process followed by an instantaneous noise process.
The second channel is called the noise channel $\Gamma_\nu$
and defined by
\begin{equation}
\Gamma_\nu [\rho]:= \Lambda_\nu [U^\dagger \rho U]
=\Lambda_\nu [(\Lambda^{(U)})^{-1}[\rho] ].
\end{equation}
Using the definition and the ideal output state
  $\rho_{out}$, it is possible to write 
the actual output state $\rho_{out}'$  such that 
\begin{equation}
\rho_{out}' = \Gamma_\nu [ U \rho_{in} U^\dagger ]
=\Gamma_\nu [\rho_{out}]. \label{noise}
\end{equation}
When the noise vanishes, the channel reduces to the identity channel:
\begin{equation}
\Gamma_0 =id_S.\label{ic}
\end{equation}
\begin{widetext}

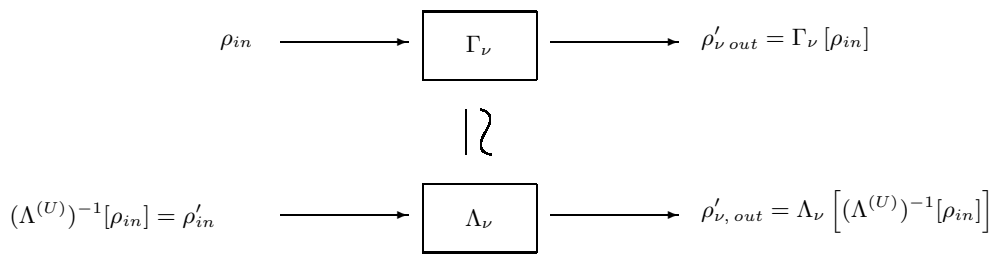
\begin{figure}[hbt]
\begin{center}
\setlength{\unitlength}{1mm}
\begin{picture}(82,35)(0,0)

\put(0,28){\makebox(5,0)[l]{$\rho_{in}$}}
\put(8,28){\vector(1,0){17}}
\put(27,23){\framebox(15,9)[c]{\shortstack[c]
{$\Gamma_{\nu}$}}}
\put(44,28){\vector(1,0){17}}
\put(64,28.5){\makebox(5,0)[l]{$\rho_{\nu\hspace{0.5mm} out}' 
= \Gamma_{\nu}\left[ \rho_{in} \right]  $}}

\put(32.7,19){\line(0,-1){6}}
\qbezier(35.2,16)(36.7,18.5)(34.6,19)
\qbezier(35.9,13)(33.7,13.5)(35.2,16)

\put(-28,5){\makebox(5,0)[l]{$( \Lambda^{(U)})^{-1} [\rho_{in} ]=\rho_{in}'$}}
\put(8,5){\vector(1,0){17}}
\put(27,0){\framebox(15,9)[c]{\shortstack[c]
{$\Lambda_{\nu}$}}}
\put(44,5){\vector(1,0){17}}
\put(64,5.5){\makebox(5,0)[l]{$\rho_{\nu, \hspace{0.5mm} out}' 
= \Lambda_\nu \left[(\Lambda^{(U)})^{-1} [\rho_{in}] \right] $}}
\end{picture}
\end{center}

\caption{{\small
In order to experimentally generate 
 the output state $\rho_{\nu, out}$ of the noise channel $\Gamma_\epsilon$ 
 for an arbitrary input state  $\rho_{in}$, 
 we take $\rho'_{in}=(\Lambda^{(U)})^{-1}[\rho_{in}]$, which is independent of $\nu$,
 as the input state for the actual channel $\Lambda_{\nu}$. }}
\end{figure}

\end{widetext}

It is stressed that despite that the noise channel $\Gamma_\nu$ is conceptual
constituent, it can be simulated in a real experiment 
by use of the actual channel $\Lambda_\nu$ (Fig.~3).
 In fact, the output state of the channel 
 $\Gamma_\nu$ defined by 
\begin{equation}
\rho_{\nu, out}=\Gamma_\nu [\rho_{in}]
\end{equation}
is exactly reproduced by
\begin{equation}
\rho_{\nu, out}= \Lambda_\nu [(\Lambda^{(U)})^{-1}[\rho_{in}]],
\end{equation}
for an arbitrary input state $\rho_{in}$. 
Therefore, by adopting a known state 
$\rho'_{in}=(\Lambda^{(U)})^{-1}[\rho_{in}] $, which is  independent of $\nu$,
 as the input state
 of the actual channel $\Lambda_\nu$, we experimentally 
obtain the output state $\rho_{\nu, out}$ of the noise channel $\Gamma_\nu$. 
 This aspect sounds very significant. 
Actually, we can replace,
  not only  theoretically but also experimentally, 
 the estimation problem for a given real 
 channel $\Lambda_\nu$ into the equivalent estimation problem for the
  noise channel $\Gamma_\nu$. 
 Hence, we later concentrate 
  on  estimation of the noise parameters for $\Gamma_\nu$ which
    satisfies  relation (\ref{ic}).

Next let us define mathematically the low-noise channel $\Gamma_\epsilon$.
 This is a kind of the noise channel and its noise parameter $\nu$ 
 takes  small positive values, which is denoted by $\epsilon$. 
We call $\epsilon$ the low-noise  parameter. 
 Physically,  $\Gamma_\epsilon$  is expected to have 
  an analytic $\epsilon$ dependence near  $\epsilon =0$. 
A rigorous mathematical formulation of this requirement is given 
 as follows.
 
Since the low-noise channel $\Gamma_{\epsilon}$ is a TPCP map,
it has a Kraus representations determined by a family of  Kraus operators.
We shall define low-noise channels in terms of their Kraus operators.
A family of TPCP maps $\Gamma_{\epsilon}$ 
with one parameter $\epsilon>0$ is called 
a {\em  low-noise channel with low-noise parameter $\epsilon$} 
if  each $\Gamma_{\epsilon}$ 
has a Kraus representation 
\begin{equation}
\Gamma_\epsilon [\rho]
=\sum_a B_a (\epsilon) \rho\, B^\dagger_{a} (\epsilon)
+\epsilon
\sum_\alpha C_\alpha (\epsilon) \rho\, C^\dagger_\alpha (\epsilon)
\label{k1}
\end{equation}
with two classes of Kraus operators $\{B_a (\epsilon)\}$ 
and $\{\sqrt{\epsilon}C_\alpha (\epsilon)\}$ 
satisfying the following conditions:
  
(i)
$B_a (\epsilon)$ is analytic at $\epsilon=0$, so that
we have the power series expansion
\begin{equation}
B_a (\epsilon) =\kappa_a {\bf 1}_S 
-\sum^\infty_{n=1} N^{(n)}_a \epsilon^n,\label{1}
\end{equation}
in a neighborhood of $\epsilon=0$,
where $\kappa_a$ and  $N^{(n)}_a$ 
are constant coefficients and  operators, respectively, 
 independent of $\epsilon$. 
The noise channel condition in \Eq{ic} requires
\begin{equation}
\sum |\kappa_a|^2 =1.
\end{equation}

(ii)
$C_\alpha (\epsilon)$ is analytic at $\epsilon=0$, so that
we have the power series expansion
\begin{equation}
C_\alpha (\epsilon) =
 M_\alpha +\sum^\infty_{n=1} M^{(n)}_\alpha \epsilon^n ,
\label{2}
\end{equation}
in a neighborhood of $\epsilon=0$,
where $M_\alpha$ and $M^{(n)}_\alpha$ are constant operators independent of 
$\epsilon$.

Needless to say, the Kraus operators satisfies the trace-preserving condition
\begin{equation}
{\bf 1}_S  =
\sum_a B^\dagger_a (\epsilon) B_a (\epsilon)
+\epsilon\sum_\alpha C^\dagger_\alpha (\epsilon) C_\alpha (\epsilon),
\label{cp}
\end{equation}
where ${\bf 1}_S$ is the identity operator.
By definition,  the relation 
\begin{equation}
\lim_{\epsilon\to +0} \Gamma_\epsilon =id_S \label{icln}
\end{equation}
is automatically satisfied. 

It should be emphasized that our definition of the low-noise channel is 
 general  from the physical point of view. Except
 that $\Gamma_\epsilon$ satisfies \Eq{icln} and
 has analytic dependence of $\epsilon $ near the origin,
 the channel $\Gamma_\epsilon$ can 
 be said to be a general quantum operation acting on the 
 input state. 
Therefore, the low-noise channel should be always 
found in the weak-interaction limit 
of $H_{SE}$ for rather general physical processes.

A useful comment is given here. Expanding \Eq{cp} in terms of $\epsilon$ 
 generates a lot of recursion relations between $\kappa_a$, $N^{(n)}_a$ and 
 $M_\alpha^{(n)}$. The higher components of the operators and the 
 coefficients  are determined recursively and systematically by solving
  the equations 
  using their lower components.
 The first-order relation in the $\epsilon$ expansion 
  of \Eq{cp} is given by
\begin{equation}
\sum_\alpha M^\dagger_\alpha M_\alpha 
=\sum_a (\kappa_a N^{(1)\dagger}_a
+\kappa_a^\ast N^{(1)}_a ). 
\label{nic}
\end{equation}

One of our fundamental interests is to ask a question:
 which input state for the low-noise channel 
 does maximize the Fisher information of its output state 
 $\rho_{\epsilon}$? By virtue of the theorem reviewed in Section 2,
 the optimal input state is a pure state.  Denote the input state by
 $|\phi\rangle\langle\phi |$. Then,
  from  \Eq{1} and \Eq{2}, 
  $\rho_\epsilon$ can be expanded as
\begin{equation}
\rho_\epsilon:=
\Gamma_{\epsilon} [|\phi\rangle\langle\phi |]
=|\phi\rangle\langle\phi | -\epsilon \rho_1 
+O(\epsilon^2).
\end{equation}
Here $\rho_1$ is given by 
\begin{eqnarray}
\rho_1 &=&
\sum_a [
\kappa_a |\phi\rangle\langle\phi|N_a^{(1)\dagger}
+
N^{(1)}_a |\phi\rangle\langle\phi|\kappa_a^\ast]
\nonumber\\
&&
-
\sum_{\alpha} M_\alpha |\phi\rangle\langle\phi |M^\dagger_\alpha.
\label{3}
\end{eqnarray}
For this output state $\rho_\epsilon$, 
we perturbatively solve the equation, 
\begin{equation}
\partial_\epsilon \rho_\epsilon = \frac{1}{2}(L_\epsilon\rho_\epsilon +\rho_\epsilon L_\epsilon),
\label{sld}
\end{equation}
in order to get the SLD operator $L_\epsilon$.
It is possible to check that the following solution actually satisfies 
 \Eq{sld} by substitution.
\begin{equation}
L_\epsilon=\frac{1}{\epsilon}
\left[
{\bf 1} -|\phi\rangle \langle \phi|
\right]
-\rho_1 +O(\epsilon).\label{4}
\end{equation}
By substituting \Eq{4} into the definition of the Fisher information,
we get the value of the information such that
\begin{equation}
J_S [ \rho_\epsilon] ={\rm Tr}[\rho_\epsilon L_\epsilon^2]
=\frac{1}{\epsilon}\langle \phi|\rho_1 |\phi\rangle +O(\epsilon^0).
\end{equation}
By using \Eq{nic} and \Eq{3}, the Fisher information is evaluated 
in the leading order of $\epsilon$ as
\begin{equation}
J_S  \left[ \rho_\epsilon \right]
=\frac{1}{\epsilon}\sum_\alpha
\left[
 \langle\phi|M^\dagger_\alpha M_\alpha|\phi\rangle
-\left|
\langle \phi|M_\alpha |\phi\rangle
\right|^2
\right]+O(\epsilon^0) .
\end{equation}

From Eq.~(\ref{Optimality}) the optimal output-measurement
observable $A_{opt}$ for any input state $\ket{\psi}$ is given by 
$A_{opt}\rho_{\epsilon}
=(J_S  \left[ \rho_\epsilon \right]^{-1}L_\epsilon+\epsilon)\rho_{\epsilon}$.
The optimal input state $|\phi_{opt}\rangle$ can be determined by maximizing
$J_S [\rho_\epsilon]$ with respect to the state $|\phi\rangle$.

Let us next discuss the low-noise channel in 
the ancilla-extended system  $S+A$. Its extended
 channel is  now given by 
 $\Gamma_\epsilon \otimes id_A$.
The input pure state $|\Psi\rangle$ 
 can be  decomposed into 
\begin{equation}
|\Psi\rangle = \sum_{n=1} C_n |n\rangle\otimes |A_n\rangle,
\end{equation}
where $\{|n\rangle\}$ is an arbitrary orthonormal basis of $S$ and 
$|A_n\rangle$'s are normalized pure states 
of $A$, which are not necessarily orthogonal to each other. 
The constants $C_n$ should satisfy the normalization condition:
\begin{equation}
\sum_n |C_n|^2 =1.
\end{equation}
The SLD for the extended state is  given by
\begin{eqnarray}
\tilde{L}_\epsilon =\frac{1}{\epsilon} (1-|\Psi\rangle\langle\Psi|) 
+O(\epsilon^0).
\end{eqnarray}
The Fisher information $J_{S+A}[\tilde{\rho}_{\epsilon}]$ for the output state 
$\tilde{\rho}_{\epsilon}=\Gamma_{\epsilon}\otimes id_{A}[\ket{\Psi}\bra{\Psi}]$
 is also evaluated in a similar manner. We have 
\begin{equation}
J_{S+A}[\tilde{\rho}_{\epsilon}]=\frac{1}{\epsilon}
\sum_\alpha
\left[
{\rm Tr}[\tilde{\rho}M^\dagger_\alpha M_\alpha]
-\left|
{\rm Tr}\left[\tilde{\rho} M_\alpha \right]
\right|^2
\right]+O(\epsilon^0),
\end{equation}
where $\tilde{\rho}$ is a state of $S$ defined by
\begin{eqnarray}
\tilde{\rho} ={\rm Tr}_{A} [|\Psi\rangle\langle\Psi|]
=\sum_{n\bar{n}}
C^\ast_{\bar{n}} \langle A_{\bar{n}}|A_n\rangle C_n |n\rangle\langle \bar{n}|.
\end{eqnarray}

The optimal output-measurement observable $\tilde{A}_{opt}$ 
for any input state $\ket{\Psi}$
is given by $\tilde{A}_{opt}\tilde{\rho}_{\epsilon}
=(J_{S+A}[\tilde{\rho}_{\epsilon}]^{-1}\tilde{L}_\epsilon
+\epsilon {\bf 1})\tilde{\rho}_{\epsilon}$.
The optimal input state $|\Psi_{opt}\rangle$ for the extended system
is determined by maximizing
$J_{S+A}[\tilde{\rho}_{\epsilon}]$ with respect to
$\tilde{\rho}={\rm Tr}_A [|\Psi\rangle\langle \Psi|]$.

If the dimension of $A$ is not less than that of $S$, we are able to 
make $|A_n\rangle$'s orthogonal to each other:
\begin{equation}
 \langle A_{\bar{n}}|A_{n}\rangle =\delta_{\bar{n} n}.
\end{equation}
Then the state $\tilde{\rho}$ is reduced into a form such that
\begin{equation}
\tilde{\rho} =\sum_n |C_n|^2 |n\rangle \langle n|.\label{5}
\end{equation}
Note that the orthonormal basis 
$\{|n\rangle \}$ of $S$ and the coefficients $C_n$ can be arbitrarily
chosen except that $\sum |C_n|^2 =1$. 
Hence, $\tilde{\rho}$ in \Eq{5} is able to describe any possible
state of $S$. 
Therefore, the dimension of the ancilla Hilbert space suffices to be
at most the same as the system Hilbert space.

 By combining both results of $J_S$ and $J_{S+A}$, we have the ancilla-assisted
  enhancement factor $\eta$ such that
\begin{equation}
\eta
=\frac{\max\left[\sum_\alpha\left[
{\rm Tr}\left[\rho_S M^\dagger_\alpha M_\alpha \right]
-
\left|
{\rm Tr}\left[\rho_S M_\alpha \right]
\right|^2\right]
\right]_{\rho_S }
}
{\max\left[\sum_{\alpha}\left[
\langle\phi_S |M^\dagger_\alpha M_\alpha|\phi_S \rangle
-
\left|
\langle \phi_S |M_\alpha |\phi_S \rangle
\right|^2\right]
\right]_{|\phi_S \rangle}
}.\label{eta}
\end{equation}
Here $\max[\ ]_{\rho_S}$ means the maximum value over all 
possible states of $S$ and 
$\max[\ ]_{|\phi_S\rangle}$ the maximum value over all
 possible pure states of $S$.

Because the set of pure states of $S$ is a subset of the set of 
 states of $S$, the following inequality  trivially holds:
\begin{equation}
\eta \geq 1.
\end{equation}

\section{Examples of Low-Noise Channels}

Low-noise channels introduced in the previous section
are found in a lot of  applications. Checking that
low-noise channels really appear in some  physical phenomena
 may lead to a deeper understanding. Thus we give two critical examples in this section.  
The details of the channels we introduce below can be seen in
 Ref.~\cite{NC00}. 

\subsection{Isotropic Depolarizing Channels}

An isotropic depolarizing channel is given by
\begin{equation}
\Gamma_\epsilon [\rho]
=
\left( 1-\frac{3}{4}\epsilon\right) \rho
+\frac{1}{4} \epsilon \sum_{a=1}^3 \sigma_a \rho \sigma_a.
\label{idc}
\end{equation}
This is a well known example
 induced by quantum noise. The parameter
 $\epsilon(\geq 0)$ is just a probability  that
 the qubit system becomes depolarized. 
The Kraus operators in \Eq{1} and \Eq{2} are given by
\begin{eqnarray}
B_0 (\epsilon) &=&\left(1-\frac{3}{4}\epsilon\right)^{1/2}{\bf 1}_S,\\
C_a (\epsilon) &=&\frac{1}{2}\sigma_a,
\end{eqnarray}
where $a=1,2,3$. 
Hence the expansion coefficients in \Eq{1} and \Eq{2} are given by
\begin{eqnarray}
\kappa_0& =&1,\\
N^{(1)}_0 &=&\frac{3}{8}{\bf 1}_S,\\
N^{(n)}_0
&=&\frac{(2n-3)!!}{n!}\left( \frac{3}{8} \right)^n {\bf 1}_S,\\
M_a &=&\frac{1}{2}\sigma_a ,\\
M_a^{(n)} &=&{\bf 0}.
\end{eqnarray}
In this case, the Fisher informations have been already calculated \cite{Fuj01}.
For the isolated original system $S$, the information is independent of the
 input state and given by    
\begin{equation}
J_S = \frac{1}{\epsilon(2-\epsilon)}.
\end{equation}
For the extended channel $\Gamma_\epsilon \otimes id_A$, the optimal input 
state is the maximally entangled state and the information is given by
\begin{equation}
\tilde{J}_{S+A} =\frac{3}{\epsilon(4-3\epsilon)},
\end{equation}
as long as the parameter $\epsilon$ is small.

\subsection{Generalized Amplitude-Damping Channels}

A generalized amplitude-damping channel is given by 
\begin{equation}
\Gamma[\rho]
=\sum^2_{a=1} B_a (\epsilon) \rho B^\dagger_a (\epsilon)
+\epsilon\sum^2_{\alpha =1} C_\alpha (\epsilon) \rho
C^\dagger_\alpha (\epsilon),\label{gadc}
\end{equation}
where $B_\nu$ and $C_\alpha$ are given by
\begin{eqnarray}
&&
B_1 (\epsilon) =
\sqrt{\frac{1}{1+e^{-\beta E}}}
\left[
\begin{array}{cc}
 1 & 0 \\
0 & \sqrt{1-\epsilon}
\end{array}
 \right],\\
&&
B_2 (\epsilon) =
\sqrt{\frac{e^{-\beta E}}{1+e^{-\beta E}}}
\left[
\begin{array}{cc}
 \sqrt{1-\epsilon} & 0 \\
0 & 1
\end{array}
 \right],\\
&&
C_1 (\epsilon) =
\sqrt{\frac{1}{1+e^{-\beta E}}}
\left[
\begin{array}{cc}
 0 & 1 \\
0 & 0
\end{array}
 \right],\\
&&C_2 (\epsilon) =
\sqrt{\frac{e^{-\beta E}}{1+e^{-\beta E}}}
\left[
\begin{array}{cc}
 0 & 0 \\
1 & 0
\end{array}
 \right].
\end{eqnarray}
The channel describes a relaxation process of the two-level system  
 driven by a finite-temperature  thermal bath. The temperature is  $(k_B \beta)^{-1}$ where $k_B$ is the Boltzmann constant.
 Here the small noise parameter $\epsilon$ is related with
 the survival rate $s$ of the initial state under 
 the relaxation such that 
 $\epsilon =1-s$. The rate $s$ is given by $s=e^{-\gamma t}$, 
  where t is time and $\gamma$ the relaxation rate constant.
 The corresponding coefficients in the $\epsilon$ expansion
  are given by  
\begin{eqnarray}
\kappa_1 &=&
\sqrt{\frac{1}{1+e^{-\beta E}}},\\
\kappa_2 &=&\sqrt{\frac{e^{-\beta E}}{1+e^{-\beta E}}},\\
N^{(n)}_1
&=&
\sqrt{\frac{1}{1+e^{-\beta E}}}\frac{(2n-3)!!}{2^n n!}
\left[
\begin{array}{cc}
 0 & 0 \\
0 & 1
\end{array}
 \right],\\
N^{(n)}_2
&=&
\sqrt{\frac{e^{-\beta E}}{1+e^{-\beta E}}}\frac{(2n-3)!!}{2^n n!}
\left[
\begin{array}{cc}
 1 & 0 \\
0 & 0
\end{array}
 \right],\\
M_1
&=&
\sqrt{\frac{1}{1+e^{-\beta E}}}
\left[
\begin{array}{cc}
 0 & 1 \\
0 & 0
\end{array}
 \right],\\
M_2
&=&
\sqrt{\frac{e^{-\beta E}}
{1+e^{-\beta E}}}
\left[
\begin{array}{cc}
 0 & 0 \\
1 & 0
\end{array}
 \right],\\
 M_{1,2}^{(n)}
&=&{\bf 0}.
\end{eqnarray}
The calculation of the Fisher information has been performed in Ref.~\cite{Fuj04}. 

The two examples in this section will be discussed again
 in Section 6.

\section{Channels on Two-Level Systems} 

In this section we concentrate  on a two-level system $S_2$ and an 
arbitrary ancilla system $A$. The dimension of $A$ is not necessarily two,
 but assumed finite.
Let us derive a universal bound on the ancilla-assisted enhancement factor $\eta$ such that
\begin{equation}
\eta \leq \frac{3}{2}.
\end{equation}
The bound must hold for all low-noise channels of $S_2$.

As well known, any state $\rho$ of the two-dimensional system $S_2$
 can be written by
\begin{equation}
\rho= \frac{1}{2}{\bf 1}_S
+\frac{1}{2} \vec{x} \cdot \vec{\sigma}
\end{equation}
where $\vec{\sigma}$ is the Pauli matrix vector and 
the three-dimensional real parameter vector $\vec{x}$ takes  values which
satisfies 
\begin{equation}
0 \leq |\vec{x}|^2 \leq 1.
\end{equation}
For pure states, the vector is normal:
\begin{equation}
|\vec{x}|^2 =1.
\end{equation}

Similarly, the matrix $M_{\alpha}$  in eqn(\ref{2}) is uniquely  expanded
as 
\begin{equation}
M_\alpha =m_{a 0}{\bf 1}_S +\sum^3_{a=1} m_{a\alpha} \sigma^a. \label{9}
\end{equation}
Now let us define complex vectors $\vec{\mu}_a$$(a=0\sim 3)$ by using the coefficients 
$m_{a\alpha}$ in \Eq{9} as
\begin{equation}
\vec{\mu}_a =(m_{a\alpha}).
\end{equation}
In the vector space, there exists a natural inner product defined by
\begin{equation}
(\vec{u},\vec{v})=\sum_{\alpha} u^\ast_\alpha v_\alpha.
\end{equation}
A metric is also induced naturally from the inner product such that 
\begin{equation}
g_{ab}:=(\vec{\mu}_a,\vec{\mu}_b )=g_{ba}^\ast ,\label{gab}
\end{equation}
where $a,b=1\sim 3$. For later convenience, define a real non-negative symmetric
 matrix $H$ by  
\begin{eqnarray}
H=[h_{ab}]=[\mbox{\rm Re } g_{ab}]\geq 0,
\end{eqnarray}
and a real three-dimensional vector $\vec{J}$ by
\begin{eqnarray}
\vec{J} =[J_a]=[\mbox{\rm Im } g_{23},\mbox{\rm  Im } g_{31}, \mbox{\rm Im } g_{12}].
\end{eqnarray}
Here denote by $h_1 ,h_2 ,h_3$ the eigenvalues of $H$.
Without loss of generality, we can assume that
\begin{equation}
0 \leq h_1 \leq h_2 \leq h_3. \label{h123}
\end{equation}

Assume later that $\vec{\mu}_a (a=1,2,3)$ are linearly independent.
Even if it is not so, because of the continuity of $\eta$, 
we can take three linearly-independent vectors $\vec{\mu}_a (t)$ parametrized
 by a real parameter $t$ such that 
\begin{equation}
\lim_{t\rightarrow 0} \vec{\mu}_a ( t)=\vec{\mu}_a.
\end{equation}
In order to get $\eta$,
 we first calculate the factor $\eta (t)$ for $\{\vec{\mu}_a (t)\}$
 and just take a limit as
\begin{equation}
\lim_{t\rightarrow 0} \eta(t) =\eta.
\end{equation}
Note that the linearly independence of $\{\vec{\mu}_a\}$ also means
\begin{equation}
H >0.
\end{equation}
This allows us to assume  the existence of $H^{-1}$.

\begin{widetext}
By a simple manipulation, we have
\begin{eqnarray}
\eta(\vec{J})
=
\frac
{
{\rm Tr} H + \vec{J} H^{-1} \vec{J}
-\min
\left[
(\vec{x} +H^{-1}\vec{J}) H (\vec{x} +H^{-1} \vec{J})
\right]_{|\vec{x}| \leq 1}
}
{
{\rm Tr} H + \vec{J} H^{-1} \vec{J}
-\min
\left[
(\vec{x} +H^{-1}\vec{J}) H (\vec{x} +H^{-1} \vec{J})
\right]_{|\vec{x}| =1}
}.\label{eta2}
\end{eqnarray}
\end{widetext}

For the original system $S$, the optimal input state
is given  by
\begin{eqnarray}
|\phi\rangle\langle \phi|= \frac{1}{2}{\bf 1}_S 
+\frac{1}{2}\vec{x}_{opt}\cdot
\vec{\sigma}
\end{eqnarray}
where  $\vec{x}_{opt}$ is the vector which minimizes
$(\vec{x} +H^{-1} \vec{J})H(\vec{x} +H^{-1}\vec{J})$ among
 whole the unit vectors.
The optimal input state $|\Psi\rangle$ for the extended system is also given 
as follows.
 Find a vector $\vec{X}$ which minimizes
$(\vec{x} +H^{-1} \vec{J})H(\vec{x} +H^{-1}\vec{J})$ among
 whole the vectors with $|\vec{x}| \leq 1$.
Then the optimal state $|\Psi\rangle$ is determined by
solving the equation
\begin{eqnarray}
Tr_A [|\Psi\rangle\langle \Psi|]
=
\frac{1}{2}{\bf 1}_S +\frac{1}{2}\vec{X}\cdot
\vec{\sigma}.
\end{eqnarray}

Let us consider the case where $\vec{J} =\vec{0}$.
The factor $\eta$ is given by
\begin{eqnarray}
\eta
&=&
\frac
{
{\rm Tr} H
-\min\left[
\sum^3_{a,b=1} x^a x^b h_{ab}
\right]_{|\vec{x}| \leq 1}
}
{
{\rm Tr} H
-\min\left[
\sum^3_{a,b=1} x^a x^b h_{ab}
\right]_{|\vec{x}| = 1}
}\nonumber
\\
&=&
\frac
{
h_{1} +h_{2}+h_{3}
}
{
h_{1} +h_{2}+h_{3}
-\min\left[
\sum^3_{a=1} h_a ({x'}^a )^2 
\right]_{|\vec{x}'| = 1}
}.\nonumber
\end{eqnarray}
Here we have made $H$ diagonalized in the last equality.
Consequently we obtain an expression of $\eta$ such that
\begin{equation}
\eta
=\frac{h_{1} +h_{2} +h_{3}}
{h_{1} +h_{2} +h_{3}-
\min\left[h_{1} ,h_{2} ,h_{3}
\right]}  .
\end{equation}
Taking account of $h_1 \leq h_2  \leq h_3$, we can easily prove 
$\eta \leq 3/2$ as follows.
\begin{eqnarray}
\eta
&=&
\frac{h_{1} +h_{2} +h_{3}}
{h_{2} +h_{3}} 
\nonumber\\
&\leq&
\frac{2h_{2} +h_{3}}
{h_{2} +h_{3}} 
\nonumber\\
&\leq&
\frac{3h_{3}}
{2h_{3}} 
=\frac{3}{2}.
\end{eqnarray}

Next let us discuss the case where  $\vec{J}\neq \vec{0}$.
 Suppose that $|H^{-1} \vec{J}|\leq 1$.
Then we have  
\begin{eqnarray}\label{eq:100}
\min
\left[
(\vec{x} +H^{-1}\vec{J}) H (\vec{x} +H^{-1} \vec{J})
\right]_{|\vec{x}| \leq 1} = 0,\label{mh}
\end{eqnarray}
because we can always  take a vector
 $\vec{x}$ such that $\vec{x} =-H^{-1}\vec{J} $. 
For later convenience, 
let us introduce a function  $G(\vec{J})$ as
\begin{eqnarray}
G(\vec{J}):= \min
\left[
(\vec{x} +H^{-1}\vec{J}) H (\vec{x} +H^{-1} \vec{J})
\right]_{|\vec{x}| = 1}. \label{Gd}
\end{eqnarray}
Then we can prove that the function $G$  satisfies 
\begin{eqnarray}
G(\vec{0}) \geq G(\vec{J}).\nonumber
\end{eqnarray}
To show this, we transform $G(\vec{J})$  as
\begin{eqnarray}
G(\vec{J})= \min
\left[
\vec{X} H \vec{X}
\right]_{|\vec{X} -H^{-1} \vec{J}| = 1}.
\end{eqnarray}
By denoting $\vec{K}=H^{-1}\vec{J}$,
 the function $G$ is given in the diagonal basis of $H$ by
\begin{equation}
G= \min
\left[
\sum_a h_a (X_a')^2
\right]_{|\vec{X}' -\vec{K}'| = 1}.
\end{equation}
Note that  the relation  $|\vec{K}'|\leq 1$ trivially holds. 
Also notice from definition (\ref{Gd}) 
that if $\vec{J}=\vec{K}'=\vec{0}$,  $G$ takes
 the minimum value of the eigenvalues of $H$, that is, $h_1$:
\begin{equation}
G(\vec{0})=h_1.
\end{equation}
To compare $G(\vec{J})$ with this value $h_1$, 
 suppose a point $\vec{X}_o'$ on a trajectory defined by
$|\vec{X}' -\vec{K}'| = 1$ such that
\begin{equation}
\vec{X}_o' =\left(K_1' \pm \sqrt{1-(K_2')^2 -(K_3')^2}, 0,0\right).
\end{equation}
Then we have
\begin{equation}
\vec{X}_o' H\vec{X}_o' 
=h_1 \left(K_1' \pm  \sqrt{1-(K_2')^2 -(K_3')^2}\right)^2.
\end{equation}
Here we fix the double sign in the above equation 
 so as to satisfy the relation:
\begin{equation}
\vec{X}_o' H\vec{X}_o' 
=h_1 \left(|K_1'| -  \sqrt{1-(K_2')^2 -(K_3')^2}\right)^2.
\end{equation}
Since the relation $|\vec{K}'|\leq 1$ holds,
 it is guaranteed that 
\begin{eqnarray}
1\geq \left(|K_1'| -  \sqrt{1-(K_2')^2 -(K_3')^2}\right)^2.
\end{eqnarray}
Therefore the important inequality
\begin{equation}
G(\vec{0}) \geq G(\vec{J}) \label{G}
\end{equation}
 really arises as follows.
\begin{eqnarray}
G(\vec{0}) &=&
h_1 
\nonumber\\
&\geq&
 h_1 \left(|K_1'| -  \sqrt{1-(K_2')^2 -(K_3')^2}\right)^2
=\vec{X}_o' H\vec{X}_o' 
\nonumber\\
&\geq& \min\left[
\vec{X'} H \vec{X'}
\right]_{|\vec{X'} -\vec{K'}| = 1} =G(\vec{J}).
\end{eqnarray}
Note that 
\begin{eqnarray}
\vec{J}H^{-1}\vec{J} \geq 0 \label{JHJ}
\end{eqnarray}
and
\begin{equation}
{\rm Tr} H -G(\vec{0})=h_2+h_3 >0. \label{h-g}
\end{equation}
Keeping \Eq{G}, \Eq{JHJ}  and \Eq{h-g} in mind, 
let us go back to the proof of $\eta(\vec{J} ) \leq \eta (\vec{0})$.
By using \Eq{mh}, we have
\begin{eqnarray}\label{eq:113}
\eta(\vec{J})
=
\frac
{
{\rm Tr} H + \vec{J} H^{-1} \vec{J}
}
{
{\rm Tr} H + \vec{J} H^{-1} \vec{J}
-G(\vec{J})}.
\end{eqnarray}
By replacing $G(\vec{J})$ by $G(\vec{0})$
in the above equality, from Eq.~(\ref{G}) we obtain
\begin{eqnarray}
\eta(\vec{J})
&\leq&
\frac
{
{\rm Tr} H + \vec{J} H^{-1} \vec{J}
}
{
[{\rm Tr} H-G(\vec{0}) ]+ \vec{J} H^{-1} \vec{J}
}.
\end{eqnarray}
By using  an inequality  such that 
\begin{eqnarray}
\frac{b+\epsilon}{a+\epsilon} \leq \frac{b}{a}
\end{eqnarray}
for $a \leq b$ and $\epsilon \geq 0$
with $\epsilon =\vec{J}H^{-1}\vec{J}$,
we have
\begin{eqnarray}
\eta(\vec{J})
&\leq&
\frac
{
{\rm Tr} H
}
{
{\rm Tr} H-G(\vec{0}) 
}=\eta (\vec{0}).
\end{eqnarray}
Consequently,  we have obtained the bound
\begin{equation}
\eta(\vec{J} ) \leq \eta (\vec{0})\leq 3/2.
\end{equation}

For the remaining case where $|H^{-1}\vec{J}| >1$, the problem becomes 
much trivial. This is because 
\begin{eqnarray}
\lefteqn{
\min
\left[
(\vec{x} +H^{-1}\vec{J}) H (\vec{x} +H^{-1} \vec{J})
\right]_{|\vec{x}| \leq 1}}\quad\nonumber\\
& =& 
\min
\left[
(\vec{x} +H^{-1}\vec{J}) H (\vec{x} +H^{-1} \vec{J})
\right]_{|\vec{x}| = 1}
\end{eqnarray}
holds in this case.
Therefore the relation $\eta =1$
 is satisfied in \Eq{eta2}.
 
Therefore, for all the possible low-noise channels, the bound
 $\eta  \leq 3/2$ 
has been proven. The equality $\eta =3/2$ can be attained by
 the channels satisfying
\begin{equation}
g_{ab}\propto \delta_{ab}
\end{equation}
 with the maximally-entangled input pure states of $S+A$. 
 
The optimal input state depends on the  vector $\vec{J}$ of the channel. 
  When $\vec{J}=\vec{0}$, the optimal input 
  state is the maximally entangled state.
If $|H^{-1} \vec{J}| \geq 1$, a factorized input state takes the maximum and
gives $\eta=1$.
 When $1> |H^{-1}\vec{J}| >0$, the optimal input state 
  is neither the maximally entangled state nor the factorized state. 
From the argument below \Eq{eq:100} the output state 
$\ket{\psi}_{S+A}$ satisfies
\begin{equation}
\mbox{\rm Tr}_A[\ket{\psi}_{S+A}\bra{\psi}_{S+A}]=\frac{1}{2}{\bf 1}_S
-\frac{1}{2}\vec{J}H^{-1}\vec{\sigma}.
\end{equation}
 The value of $\eta$ given by \Eq{eq:113}
also changes continuously between
 $1\leq \eta <3/2$ depending on $|H^{-1}\vec{J}|$.
 
 The channel dependence of the optimal input state has been already 
 noticed in a generalized amplitude-damping channel 
 \cite{Fuj04} by changing the temperature of the thermal bath. 
 Because of the simplicity of the model, it is possible to 
 estimate the unknown parameter even in a finite parameter region. 
 On the other hand, in this paper, 
  the parameter region of the low-noise channel is constrained
  to a neighborhood of a fixed value ($\epsilon =0$).
  However, we would like to stress that our channel
  includes an enormous number of degrees of freedom corresponding to
 $\kappa_a$, $N^{(n)}_a$ and $M^{(n)}_\alpha$, compared with
  the generalized amplitude-damping channel. 
 
Note that the isotropic depolarizing channel  (\Eq{idc} in Section 5) 
 is one of the 
channels attaining the bound ($\eta =3/2$). 
The vectors $\vec{\mu}_a$ are calculated as
\begin{eqnarray}
&&
\vec{\mu}_1
=\frac{1}{2}
\left[
\begin{array}{c}
1 \\
0 \\
0
\end{array}
\right],\\
&&
\vec{\mu}_2
=\frac{1}{2}
\left[
\begin{array}{c}
0 \\
1 \\
0
\end{array}
\right],\\
&&
\vec{\mu}_3
=\frac{1}{2}
\left[
\begin{array}{c}
0 \\
0 \\
1
\end{array}
\right].
\end{eqnarray}
The corresponding matrix $g_{ab}$ is just evaluated as
\begin{equation}
g_{ab} =\frac{1}{4}\delta_{ab}.
\end{equation}
 Thus the channel can achieve $\eta =3/2$.
 
On the other hand, 
the generalized amplitude-damping
 channels \Eq{gadc} in Section 5) cannot achieve the bound.
 The vectors $\vec{\mu}_a$ are now described by
\begin{eqnarray}
&&
\vec{\mu}_1
=\frac{1}{2}
\left[
\begin{array}{c}
\sqrt{\frac{1}{1+e^{-\beta E}}} \\
\sqrt{\frac{e^{-\beta E}}{1+e^{-\beta E}}} \\
0
\end{array}
\right],\\
&&
\vec{\mu}_2
=\frac{i}{2}
\left[
\begin{array}{c}
\sqrt{\frac{1}{1+e^{-\beta E}}} \\
-\sqrt{\frac{e^{-\beta E}}{1+e^{-\beta E}}} \\
0
\end{array}
\right],\\
&&
\vec{\mu}_3
=
\left[
\begin{array}{c}
0 \\
0 \\
0
\end{array}
\right].
\end{eqnarray}
The corresponding $g_{ab}$ is now given by 
\begin{eqnarray}
[g_{ab}]
=
\frac{1}{4}
\left[
\begin{array}{ccc}
1 & i\frac{1-e^{-\beta E}}{1+e^{-\beta E}} &0 \\
- i\frac{1-e^{-\beta E}}{1+e^{-\beta E}} & 1 & 0\\
0 &0&0
\end{array}
\right].
\end{eqnarray}
Because $g_{ab} \propto \delta_{ab}$ does hold, 
the channel  cannot satisfy $\eta =3/2$ for any parameter value.
 In spite of the ancilla extension,
  the ancilla-assisted  enhancement does not appear at all ($\eta =1$), 
as long as the low-noise parameter $\epsilon$ is small enough. 
 This is because the value of $|H^{-1}\vec{J}|$ 
 diverges and the relation $|H^{-1}\vec{J}| >1$ always holds.


\begin{acknowledgments}
The authors thank Akio Fujiwara and Gen Kimura for useful comments
and discussions.
This work was supported by the SCOPE project of the MPHPT of Japan
and by the Grant-in-Aid for Scientific Research of the JSPS.
\end{acknowledgments}



\end{document}